# Thickness Dependence of the Physical Properties of Atomic-Layer Deposited Al$_2$O$_3$

Yael Etinger-Geller[1], Ekaterina Zoubenko[1], Maria Baskin[2], Lior Kornblum[2]* and Boaz Pokroy[1,3]*

[1]*Department of Materials Science and Engineering, Technion – Israel Institute of Technology, Haifa 32000, Israel*

[2]*Andrew & Erna Viterbi Department of Electrical Engineering, Technion – Israel Institute of Technology, Haifa 32000, Israel*

[3]*Russell Berrie Nanotechnology Institute, Technion – Israel Institute of Technology, Haifa 32000, Israel*

Inspired by nature, we investigate the short-range order effect on the physical properties of amorphous materials. Amorphous Al$_2$O$_3$ thin films exhibit a higher proportion of their 4-coordinated Al sites close to the surface, causing variations in the average short-range order of the film. Below some thickness, the density of these films changes with size. In this work, we address the short-range order effect, through the thickness, on the electronic and optical properties of atomic layer deposited (ALD) Al$_2$O$_3$ thin films. Both the refractive index and the permittivity were found to vary with size. The refractive index increased with thickness, and for thick films (~50 nm) it was comparable to that of bulk amorphous Al$_2$O$_3$. The permittivity increased with thickness as well, but did not attain those of the bulk material. We discuss how these effects correlate with the density and short-range order. These results shed light on the size effects in thin amorphous oxides, and may guide the design of electronic and optical components and devices.

## I.     Introduction

Amorphous phases are commonly found in nature when controlled mineralization is required; they enable the organism to exert control over the crystallization, thereby achieving different structures and morphologies according to necessity[1-3]. To gain this control, the organism manipulates the short-range order within the amorphous phase[2]. Whereas in nature such control is achieved through different additives[1,2], in amorphous thin films the short-range order was found to vary with size[4,5]. Amorphous Al$_2$O$_3$ thin films deposited by atomic layer deposition (ALD) have shown size-dependent structural variations, where thinner films exhibited a higher proportion of their 4-coordinated Al (Al$_4$) sites than thicker ones. It was also shown that the density of the films increases with size by more than 15%.[5] Since several physical properties of a material are density-dependent,[6,7,8] we expect that technologically important electronic and optical properties would exhibit significant variations with thickness.

We test this hypothesis on films grown by ALD, which allows fabrication of amorphous oxides of excellent quality, with precise and conformal morphology[9,10,11,12]. ALD $Al_2O_3$ is one of the most common amorphous oxides in use for science and technology, owing to its large growth window and useful optical and electronic properties. As such, thin films of $Al_2O_3$ are useful for antireflective coatings[13], in particular the quarter wavelength type used in optical sensors[14], for MEMS applications[15] and other.

Motivated by thickness-density relations and its potential effects on the physical properties, we focus this work on the electronic and optical properties of the technologically important ALD-$Al_2O_3$ system. Demonstrating a new approach for tuning these properties is expected to contribute to design and application of these materials in optical and electronic components and devices.

## II. Experimental

Deposition of amorphous $Al_2O_3$ by ALD is a common practice and has been well studied[10,11,16,17]. Given its very wide ALD window this procedure serves as an ALD model process, and the resulting thin films are of high quality, smooth and pin-hole free[9,10]. $Al_2O_3$ films were grown in a plasma-enhanced ALD reactor (PEALD, ALD R-200 Advanced, Picosun, Finland) at 200°C using trimethyl aluminum and water, as described elsewhere[5]. The thickness of the films was measured with x-ray reflectivity (XRR, SmartLab, Rigaku, Japan) and spectroscopic ellipsometry (VASE, Woollam, USA).

Optical characterization was conducted on films deposited on p-Si (100) wafers. Spectroscopic ellipsometry was performed in the wavelength range of 300−1000 nm at three different angles (65, 70 and 75 degrees). An interfacial oxide layer was accounted for in the interpretation of the results. Metal-oxide-semiconductor (MOS) capacitors were fabricated on n-type Si wafers with 11 nm of $SiO_2$ grown by dry thermal oxidation. ALD $Al_2O_3$ films were deposited directly onto the $SiO_2$ layer. 50 nm thick Al pads were deposited using e-beam evaporation (Airco Temescal FC-1800) through a shadow mask, and 300 nm blanket Al was deposited for a back contact.
Capacitance−voltage (C−V) measurements were performed using an Agilent E4980A LCR meter (Agilent Technologies, USA), with the capacitance corrected for series resistance based on multi-frequency analysis.[18]

## III. Results

The optical properties of amorphous materials make these materials attractive for various applications in science and technology[19]. The nature of the relationship between a material's density and its refractive index has prompted different theories, in all of which it is assumed that an increase in density should result in a higher refractive index. To study the refractive index of the $Al_2O_3$ layers we used spectroscopic ellipsometry; thin $Al_2O_3$ films of various thicknesses ranging between 15 and 65 nm were deposited directly onto Si wafers and scanned with a spectroscopic ellipsometer. The results were fitted using the "VAZE 32" software. Changes in the refractive index with different $Al_2O_3$ thicknesses can be seen in FIG. 1a:

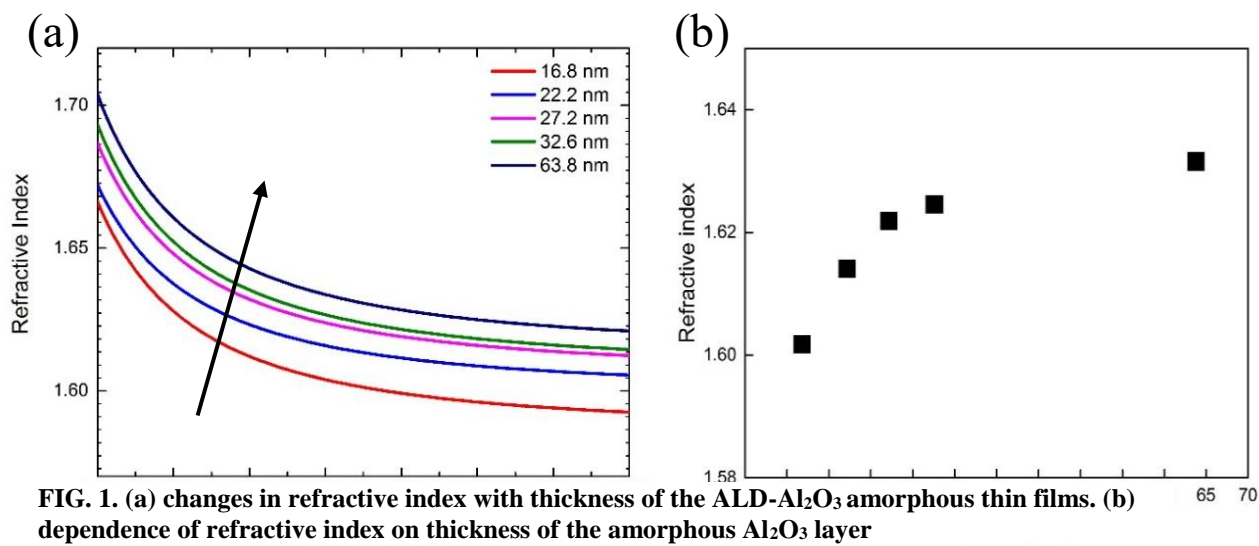

FIG. 1. (a) changes in refractive index with thickness of the ALD-$Al_2O_3$ amorphous thin films. (b) dependence of refractive index on thickness of the amorphous $Al_2O_3$ layer

The figure shows that as the thickness increases, so does the refractive index. This result coincides with previous findings indicating that density increases with thickness, since a change in density should result in a corresponding change in the refractive index.

Since the value of the refractive index does not change significantly (<1%) at wavelengths higher than 600 nm, it can be considered to be constant above that level. The selected value of refractive index was at wavelength of 635 nm (comparable with results from other studies, achieved with a single-wavelength ellipsometer, at wavelength of 632.8 nm). In FIG. 1b, the relationship between refractive index and thickness is plotted; As shown, for thicknesses below 40 nm the refractive index gradually increases until it reaches a constant value of ~1.63. This value is close to values accepted for amorphous aluminum oxide in previous studies[16,20]. This relationship is reminiscent of the relationship, reported in the past[5], between the refractive index and the thickness. Thus, as expected from both the theory and previous findings, the refractive index increases with density.

Another important parameter that is theoretically affected by the density is the dielectric constant (k). This parameter was studied using MOS capacitors with varying thicknesses of $Al_2O_3$. C−V analysis at 1 MHz was employed to determine the dielectric constant (FIG. 2a), accounting for the bottom 11 nm $SiO_2$ layer, which was uniform across samples.

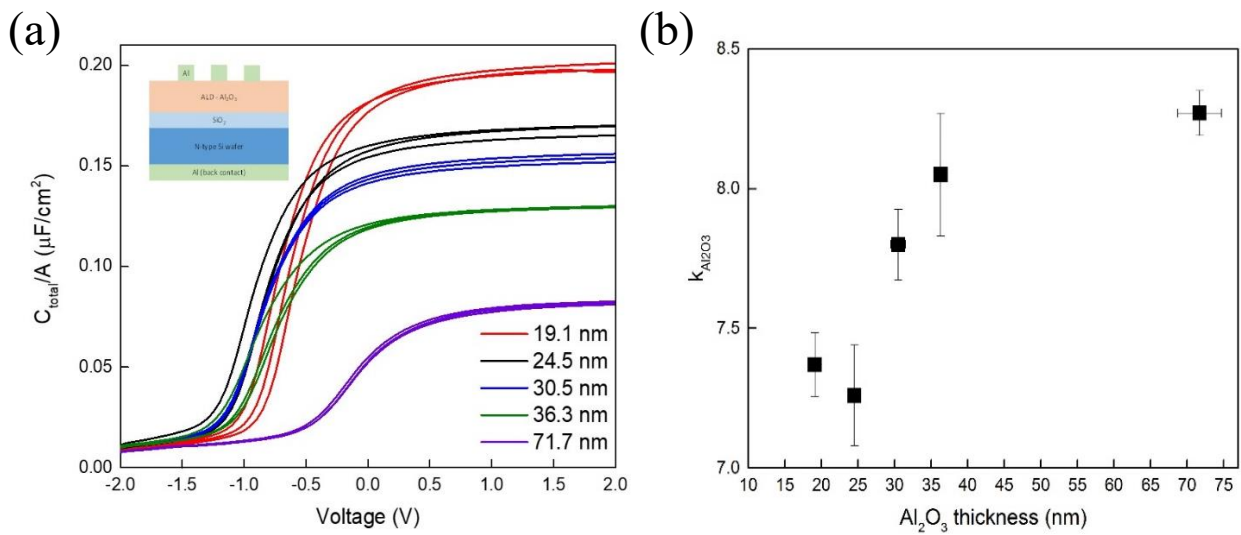

**FIG. 2. (a) C−V curves achieved for the different capacitors, with different thicknesses of $Al_2O_3$ layer, normalized to the measured contact area measured at 1 MHz; inset: Schematic description of the MOS capacitors used to determine the dielectric constants of $Al_2O_3$ films (b) Dependence of the dielectric constant on thickness of the $Al_2O_3$ layer**

The C-V curves (FIG. 2a) exhibit well-behaved characteristics[21] that indicate negligible (and uniform) contribution of interface states. The maximum value of each curve (accumulation, in positive voltages) was used to obtain the capacitance of the oxides.

A summary of the resulting dielectric constants (FIG. 2b) reveals that the dielectric constant increases with increasing thickness until it reaches a value of ~8.3, which is lower than the dielectric constant of crystalline $Al_2O_3$ phases.[22] It is also known that crystallinity increases its value[23]. It can be seen that the dielectric constant at 24.5 nm is slightly lower than that of the 19.1 nm thick sample; however the difference between the two values is within the error range of the results. Since, the density also increases with thickness and, according to these results, thicker films exhibited higher dielectric constants, these findings agree with our prediction. Therefore, the thickness can be used, in some cases, as a tuning parameter for the dielectric constant.

## IV. Discussion

### A. Dependence of the Refractive Index on Thickness

The refractive index of a thin $Al_2O_3$ film was found to vary with thickness, with higher thicknesses yielding higher refractive indexes. For thick films, the accepted refractive index is similar to that of bulk amorphous $Al_2O_3$ (~1.63), but the values obtained for thin films in the present study were lower, indicating that by changing the thin film's thickness it is possible to manipulate this optical property according to a specific requirement.

In our previous study we examined the relationship between the density of an amorphous $Al_2O_3$ film and its critical angle $(\theta_c)$[5]. According to Snell's law, the ratio between the refractive indexes (n) and the refraction angles ($\theta$) when a ray passes from one medium to another is defined by Eq. 4[24]:

$$\frac{n_1}{n_2} = \frac{\sin(\theta_2)}{\sin(\theta_1)} \quad (1)$$

In our case, assuming that the x-ray passes from air ($n_1 = 1$) to our layer and that total reflection occurs ($\theta_2 = 90°$), the relationship between the critical angle and the refractive index of the layer would be:

$$n = \sin(\theta_C) \quad (2)$$

This relationship can be plotted (FIG. 3), since the critical angle for total reflection can be found by XRR, as reported previously[5,25].

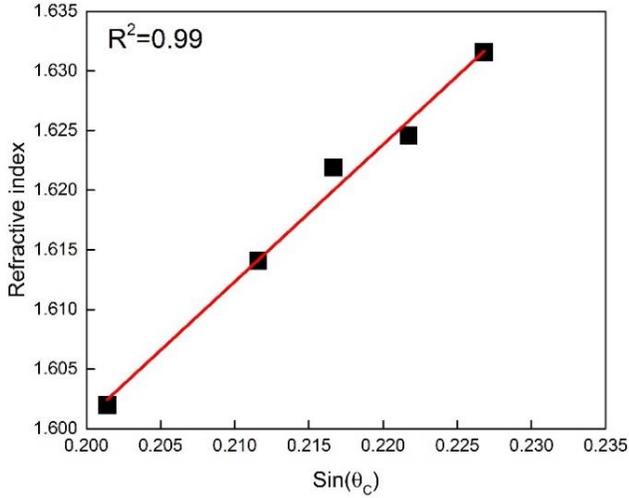

**FIG. 3. Relationship between the critical angle (measured by XRR) and the refractive index (measured by ellipsometry)**

Good linearity is observed between the two values, with $R^2 = 0.99$. The proportionality factor is 1.15, which is close to 1, the theoretical factor, supporting the high correlation between the measured parameters.

Another parameter that changes with thickness, as previously shown, is the density. The general relationship between the density ($\rho$) and the refractive index (n), according to Anderson and Schreiber, which takes into consideration an overlap field for near-neighbor interaction is the following[8]:

$$\frac{n^2 - 1}{\left[4\pi + b(n^2 - 1)\right]\rho} = \frac{\alpha}{M} \equiv R \quad (3)$$

where b is the electronic overlap parameter, which is unique to each material and can be found via the extrapolation of n-ρ data, M is the molecular weight and α is the polarizability, which is the ability to form a dipole. It is also possible to rewrite Eq. (3) as follows:

$$\frac{4\pi}{n^2-1} = \frac{1}{R} \cdot \frac{1}{\rho} - b \qquad (4)$$

Our previous findings, which concern the relationship between thickness and density[5], enable us to plot Eq. (4) as shown in FIG. 4:

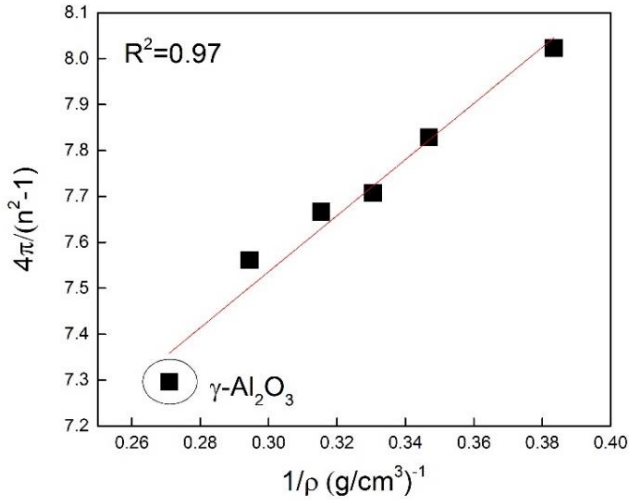

**FIG. 4. Dependence of the refractive index on density, according to Anderson and Schreiber**

Since the amorphous $Al_2O_3$ ALD films exhibit similar structural features to those of $\gamma$-$Al_2O_3$[4,5], the theoretical parameters for $\gamma$-$Al_2O_3$[26], have been added to the figure. These values can be seen to show good linearity, with $R^2 = 0.97$, further validating the correlation between the density (owing to thickness variations) and the refractive index in amorphous ALD thin films of $Al_2O_3$. By subjecting the results to linear regression, it is possible to extract that b = −5.7 ± 0.2 and R = 0.16 ± 0.02.

**B. Dependence of the Dielectric Constant on Thickness**

According to our present results, it is clear that the dielectric constant increases with thickness of the amorphous film. It was previously shown that density also changes with thickness[5], implying that the dielectric constant changes due to density variations.
The values obtained in our experiments are similar to values achieved in other studies[16,23,27]. It was shown in the past that the dielectric constant of amorphous $Al_2O_3$ changes with the thickness, but the change was attributed to other factors, such as interfacial layers or quantum effects[27,28]. Our use of 11 nm thermally-grown $SiO_2$ layer and its separate analysis in the present work, precludes contributions of interfacial effects that may affect the results.

Many studies have been carried out on various materials, in which the change in dielectric constant was studied as a function of density, and it was found that higher density yields higher dielectric constant[7,29].

The relationship between the density and the dielectric constant is plotted Fig. 8. Theoretical values for crystalline γ- $Al_2O_3$ were added as well[30]

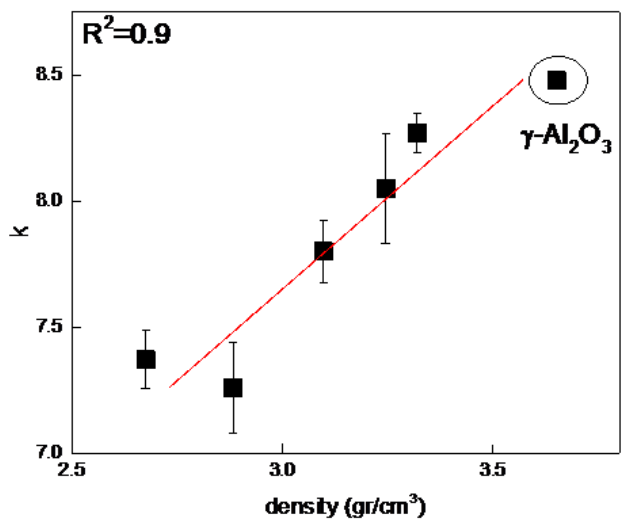

**FIG. 5. Dependence of dielectric constant on density**

It can be seen that the values align well, indicating on a clear correlation between the density and the dielectric constant. This is an important finding, as it can open a variety of new possibilities for amorphous dielectric materials.

## V. Summary and Conclusions

Amorphous ALD $Al_2O_3$ films are commonly used and are of significant interest for various applications. Previous studies showed that the short-range order close to the surface in these films differs from that in the bulk amorphous $Al_2O_3$; hence, the average short-range order in thinner films differs from that in thicker ones. These variations were previously found to affect the density of the films, pointing to a potential for tuning density-dependent properties. It is shown here that the refractive index and the dielectric constant of ALD $Al_2O_3$ films change with size, owing to variations in film density: thinner films, which have lower density, exhibit lower refractive indices and dielectric constants than those of the higher-density, thicker films. This finding implies the possibility of tuning these properties solely by size. This effect is not expected to be limited $Al_2O_3$, and should manifest in other amorphous systems, where it might further emerge at different thicknesses.

**Acknowledgements**

The research leading to these results received funding from the European Research Council under the European Union's Seventh Framework Program (FP/2013-2018) / ERC Grant Agreement no. 336077. We are grateful to Dr. Oleg Kreinin for his help with preparing the samples and for operating the ALD system. We are also grateful to Dr. Guy Ankonina for his help in modeling the results obtained using the spectroscopic Ellipsometer. Sample fabrication was done with the support of the Technion's Micro-Nano Fabrication unit (MNFU). We thank Tomer Stein and David Shapiro for helping with the analysis. L.K. is a Chanin Fellow 1 Laurie B Gower, Chemical reviews **108** (11), 4551 (2008).
2 Lia Addadi, Sefi Raz, and Steve Weiner, Advanced Materials **15** (12), 959 (2003).
3 Joanna Aizenberg, Lia Addadi, Stephen Weiner, and Gretchen Lambert, Advanced Materials **8** (3), 222 (1996).
4 Leonid Bloch, Yaron Kauffmann, and Boaz Pokroy, Crystal Growth & Design **14** (8), 3983 (2014).
5 Yael Etinger-Geller, Alex Katsman, and Boaz Pokroy, Chemistry of Materials **29** (11), 4912 (2017).
6 Sharmila Devi and Daryl R Williams, European Journal of Pharmaceutics and Biopharmaceutics **88** (2), 492 (2014);   AA Louis, Journal of Physics: Condensed Matter **14** (40), 9187 (2002);   Stuart O Nelson, Transactions of the ASAE **26** (6), 1823 (1983).
7 Yang Zhang, Jichu Yang, and Yang-Xin Yu, The Journal of Physical Chemistry B **109** (27), 13375 (2005).
8 B Marler, Physics and chemistry of minerals **16** (3), 286 (1988).
9 Ville Miikkulainen, Markku Leskelä, Mikko Ritala, and Riikka L Puurunen, Journal of Applied Physics **113** (2), 021301 (2013).
10 Markku Leskelä and Mikko Ritala, Thin solid films **409** (1), 138 (2002);   Steven M George, Chemical reviews **110** (1), 111 (2009).
11 Richard W Johnson, Adam Hultqvist, and Stacey F Bent, Materials Today **17** (5), 236 (2014).
12 Markku Leskelä and Mikko Ritala, Angewandte Chemie International Edition **42** (45), 5548 (2003).
13 Adriana Szeghalmi, Michael Helgert, Robert Brunner, Frank Heyroth, Ulrich Gösele, and Mato Knez, Applied Optics **48** (9), 1727 (2009);   Roland Thielsch, Alexandre Gatto, and Norbert Kaiser, Applied optics **41** (16), 3211 (2002).
14 Ying Zhao, Fufei Pang, Yanhua Dong, Jianxiang Wen, Zhenyi Chen, and Tingyun Wang, Optics Express **21** (22), 26136 (2013).
15 Steven M George and Cari F Herrmann, (Google Patents, 2009);   O. Hahtela, P. Sievilä, N. Chekurov, and I. Tittonen, Journal of Micromechanics and Microengineering **17** (4), 737 (2007).
16 MD Groner, FH Fabreguette, JW Elam, and SM George, Chemistry of Materials **16** (4), 639 (2004).
17 SD Elliott, G Scarel, C Wiemer, M Fanciulli, and G Pavia, Chemistry of materials **18** (16), 3764 (2006).
18 Edward H Nicollian, John R Brews, and Edward H Nicollian, *MOS (metal oxide semiconductor) physics and technology*. (Wiley New York et al., 1982).

19. Akihisa Inoue and Koji Hashimoto, *Amorphous and nanocrystalline materials: Preparation, properties, and applications*. (Springer Science & Business Media, 2013); Zbigniew H Stachurski, Materials **4** (9), 1564 (2011).
20. Zi-Yi Wang, Rong-Jun Zhang, Hong-Liang Lu, Xin Chen, Yan Sun, Yun Zhang, Yan-Feng Wei, Ji-Ping Xu, Song-You Wang, and Yu-Xiang Zheng, Nanoscale research letters **10** (1), 46 (2015); A Ortiz, JC Alonso, V Pankov, A Huanosta, and E Andrade, Thin Solid Films **368** (1), 74 (2000).
21. Lior Kornblum, Jonathan A. Rothschild, Yaron Kauffmann, Reuven Brener, and Moshe Eizenberg, Physical Review B **84** (15), 155317 (2011).
22. Kenneth L Carr, (Google Patents, 2013).
23. Stefan Jakschik, Uwe Schroeder, Thomas Hecht, Martin Gutsche, Harald Seidl, and Johann W Bartha, Thin Solid Films **425** (1-2), 216 (2003).
24. David T Blackstock, *Fundamentals of physical acoustics*. (John Wiley & Sons, 2000).
25. Isao Kojima and Boquan Li, The Rigaku Journal **16** (2), 31 (1999).
26. Atul Khanna and Deepak G Bhat, Surface and Coatings Technology **201** (1-2), 168 (2006); R Nakamura, T Shudo, A Hirata, M Ishimaru, and H Nakajima, Scripta Materialia **64** (2), 197 (2011).
27. MD Groner, JW Elam, FH Fabreguette, and Sm M George, Thin Solid Films **413** (1-2), 186 (2002).
28. S-H Lo, Douglas A Buchanan, and Yuan Taur, IBM Journal of Research and Development **43** (3), 327 (1999).
29. VL Lanza and DB Herrmann, Journal of Polymer Science Part A: Polymer Chemistry **28** (118), 622 (1958); Stuart O Nelson, IEEE Transactions on Instrumentation and Measurement **54** (5), 2033 (2005).
30. Choong-Ki Lee, Eunae Cho, Hyo-Sug Lee, Kwang Soo Seol, and Seungwu Han, Physical Review B **76** (24), 245110 (2007).